# Versatile Graphene-Based Platform for Robust Nanobiohybrid Interfaces

Rebeca Bueno,[†] Marzia Marciello,[†,‡] Miguel Moreno,[§] Carlos Sánchez-Sánchez,[†] José I. Martinez,[†] Lidia Martinez,[†] Elisabet Prats-Alfonso,[∥,⊥] Anton Guimerà-Brunet,[∥,⊥] Jose A. Garrido,[#] Rosa Villa,[∥,⊥] Federico Mompean,[†] Mar García-Hernandez,[†] Yves Huttel,[†] María del Puerto Morales,[†] Carlos Briones,[§] María F. López,[†] Gary J. Ellis,[¶] Luis Vázquez,[†] and José A. Martín-Gago*,[†]

[†]Materials Science Factory, Institute of Materials Science of Madrid (ICMM-CSIC), Sor Juana Inés de la Cruz 3, 28049 Madrid, Spain

[‡]Nanobiotechnology for Life Sciences Group, Department of Chemistry in Pharmaceutical Sciences, Faculty of Pharmacy, Complutense University (UCM), Plaza Ramón y Cajal s/n, 28040 Madrid, Spain

[§]Laboratory of Molecular Evolution, Centro de Astrobiología (CSIC-INTA), Torrejón de Ardoz, 28850 Madrid, Spain

[∥]Instituto de Microelectrónica de Barcelona IMB-CNM (CSIC) Esfera UAB, Bellaterra, 08193 Barcelona, Spain

[⊥]Centro de Investigación Biomédica en Red en Bioingeniería Biomateriales y Nanomedicina (CIBER-BBN), 28029 Madrid, Spain

[#]Catalan Institute of Nanoscience and Nanotechnology (ICN2) CSIC and The Barcelona Institute of Science and Technology Campus UAB, Bellaterra, 08193 Barcelona, Spain

[¶]Polymer Physics Group, Institute of Polymer Science and Technology (ICTP-CSIC), Juan de la Cierva 3, 28006 Madrid, Spain

**ABSTRACT:** Technologically useful and robust graphene-based interfaces for devices require the introduction of highly selective, stable, and covalently bonded functionalities on the graphene surface, whilst essentially retaining the electronic properties of the pristine layer. This work demonstrates that highly controlled, ultrahigh vacuum covalent chemical functionalization of graphene sheets with a thiol-terminated molecule provides a robust and tunable platform for the development of hybrid nanostructures in different environments. We employ this facile strategy to covalently couple two representative systems of broad interest: metal nanoparticles, via S−metal bonds, and thiol-modified DNA aptamers, via disulfide bridges. Both systems, which have been characterized by a multitechnique approach, remain firmly anchored to the graphene surface even after several washing cycles. Atomic force microscopy images demonstrate that the conjugated aptamer retains the functionality required to recognize a target protein. This methodology opens a new route to the integration of high-quality graphene layers into diverse technological platforms, including plasmonics, optoelectronics, or biosensing. With respect to the latter, the viability of a thiol-functionalized chemical vapor deposition graphene-based solution-gated field-effect transistor array was assessed.

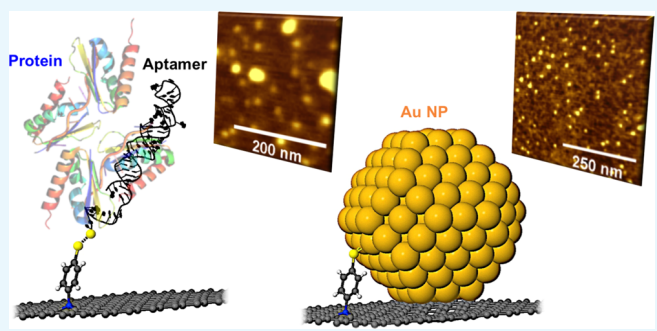

## 1. INTRODUCTION

Whilst pristine graphene is one of the most relevant materials of the decade, several important shortcomings must be overcome before it may step from fundamental physics to applied technology.[1] In particular, the absence of an electronic band gap and its extreme chemical inertness undoubtedly compromise its use as an active element in electronic devices or hybrid structures. Molecular functionalization of graphene can provide singular and advantageous properties, and there have been many attempts via nondestructive methodologies to furnish graphene with surface modifications whilst attempting to preserve its extraordinary properties.[2−5] Evidently for technological applications in environmental operating conditions, such as those related to biosensing, strong and stable molecular links are required, preferably via covalent functionalization. The most common covalent functionalization methodologies for graphene are chemical routes,[6−8] mainly based on the reaction between free radicals or dienophiles and the C═C bonds of pristine graphene.[9] However, although well-developed wet chemistry routes may succeed to link many interesting functional groups to the graphene surface, they usually fall short in their usefulness either due to a low degree of functionalization or to extreme disruption of the surface due to the harsh nature of the reaction conditions,[9,10] resulting in graphene platforms where excessive defect concentrations degrade the outstanding properties of graphene, thus limiting its applicability.

Here, we use a new, recently reported strategy[11] for the selective functionalization of graphene based on the controlled formation of atomic vacancies in order to obtain a uniformly covered surface with a covalently bound spacer molecule that is formed from the spontaneous bonding of p-aminothiophenol (pATP) molecules at the vacancies. This results in the controlled decoration of the graphene surface with active thiol moieties that can be directly used to bond diverse nanoarchitectures to graphene. We show that although the functionalization protocol is undertaken in ultrahigh vacuum (UHV), the thiol functional moiety is robust and stable in different environments. As a consequence, it can be used, for example, for the immobilization of metal nanoparticles (NPs), particularly gold NPs (Au NPs) which are known to show a high affinity toward the thiol group.[12] The deposition of NPs on graphene sheets has become a valuable strategy for coupling graphene with plasmonic nanostructures[13] and shows promise for optoelectronic materials[14,15] or in (bio)sensing[16] or energy storage[17] applications. The initial graphene substrate generally employed is graphene oxide (GO), which although it is significantly more reactive than pristine graphene, allowing the chemical binding of the NPs to the surface via reduction of the GO and the metal salts, this is at the expense of significantly degraded electrical and electronic properties. Although recently a nonchemical Au NP decoration method using laser ablation in liquids was reported,[18] a GO substrate was still required to efficiently bind the NPs to the surface.

On the other hand, thiol chemistry is an ideal tool to couple a wide range of molecular architectures, in particular biomolecules through the formation of strong sulfide bridges.[19] One example is nucleic acid aptamers,[20] which comprise RNA or single-stranded DNA (ssDNA) oligonucleotides selected in vitro from a vast library of synthetic random oligonucleotides[21] that can bind with high affinity and specificity to a given target molecule. Aptamer-based biosensors have recently emerged as improved biorecognition elements and are increasingly used in biotechnology, biomedicine, and environmental control.[22,23] Moreover, several are the advantages of using graphene-based platforms for biointerfaces,[24,25] among them low noise, flexibility, and more interestingly, that because of the electronic nature of graphene, sensing and actuator materials are on the same device. This work opens the door toward the integration of graphene as a sensing platform.

Finally, it is important to remark that whilst the UHV environment provides highly controlled preparation of functional material surfaces, these generally fail when exposed to the atmosphere. The 13 orders of magnitude difference in pressure, commonly known as the pressure-gap that separates the real world from the artificially defined world of surface science, is a very difficult barrier to overcome. In this work, both worlds are successfully combined. UHV protocols and characterization techniques for covalent attachment and characterization of the process are employed, and solution-based and atmospheric protocols are used to incorporate technologically important nanoarchitectures (see Figure 1). These conjugated NPs and biomolecules are representative examples of nanobiohybrid nanostructures that may provide excellent platforms with wide applicability in diverse scientific and technological fields.[26]

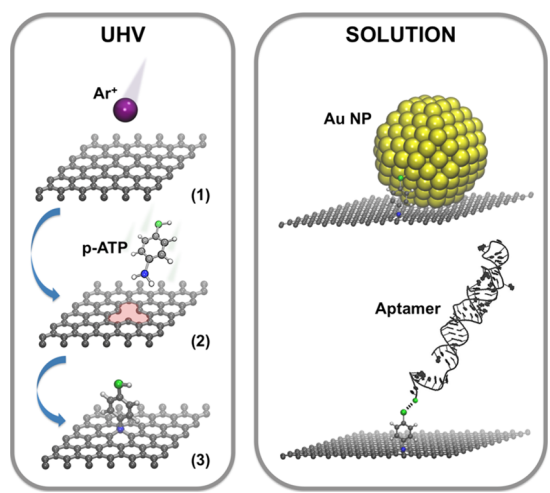

**Figure 1.** Schematic illustration of the two-step functionalization process: Left panel: UHV linking of the aminothiophenol molecules. (1) Irradiation of the graphene surface with Ar+ at 140 eV; (2) exposure of the graphene surface with ion-beam-induced vacancies to pATP molecules; and (3) covalent bonding of N of pATP to the carbon network at the induced vacancy, leaving the thiol group exposed to the medium. Right panel: in-solution process of the thiol group. Top: scheme of the thiol linkage to the Au NPs; bottom: conjugation of the thiol-modified ssDNA aptamer to the thiol group through a disulfide bond formation.

## 2. RESULTS AND DISCUSSION

The mechanism mediating the functionalization of graphene has been understood in terms of the extra charge accumulated in the broken dangling bonds around the single atomic vacancies that selectively oxidize any molecule containing an amino moiety (Figure 1, step1). When pATP molecules are injected into the chamber in the gas phase (Figure 1, step 2), the N atom from the primary amino group is incorporated in the graphene network (Figure 1, step 3). The nature of the molecular bonding to the surface is covalent, as confirmed by the S 2p and N 1s X-ray photoelectron spectroscopy (XPS) core-level peaks (Figure 2b) and density functional theory (DFT) calculations. The N 1s core level peak can be fitted with a single component at 400.09 eV, in good agreement with the values reported for N within the graphene network and with a radical.[11,27] The calculations show that the molecule strongly binds within the surface lattice with an anchoring energy (energy gain due to the saturation of the C dangling bonds) of 12.21 eV after the reaction is produced, yielding a highly stable final structural configuration (depicted in step 3 of Figure 1). The result of this process is that the free thiol group of the pATP is exposed to the medium. This is opposite to the process that occurs in thiol−ene click chemistry routes, where the thiol group binds to the pristine graphene surface.[28−30] In our case, the wide and rich chemistry of the thiol groups can be used to further attach other relevant structures and molecules to the graphene sheet, either via S−metal bonds or by the formation of disulfide bridges.[20]

The existence of a stable thiol group after steps 1−3 has been verified by XPS spectra recorded on the same experimental system where the functionalization was performed that means without exposure to air. The S 2p core level peak, shown in Figure 2a, appears at 163.9 eV, confirming the presence of the thiol moiety, which is easily differentiated from S dimers, sulfur oxides, or other S-containing species. The XPS spectrum can be fitted with a unique component with a spin-orbit splitting of 1.15 eV, indicating that all of the molecules are at the same chemical state. Furthermore, these molecules were observed by scanning tunneling microscopy (STM) (on the same system where functionalization was performed) to be well dispersed over the surface, Figure 2c, where the molecules appear as bright bumps in the images with typical apparent heights of 2.1 ± 0.5 Å (statistics over 150 protrusions), a value similar to that reported for aminophenol.[11]

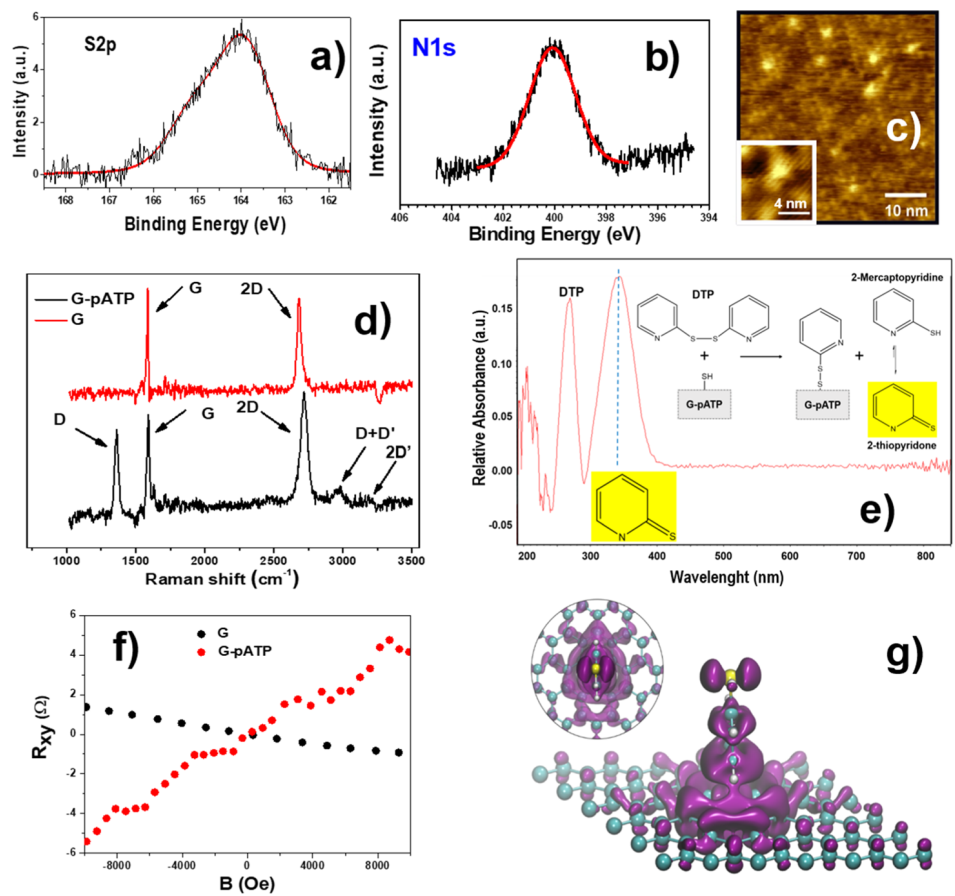

**Figure 2.** (a) S 2p XPS core-level peak of the G-pATP surface corresponding to free thiol groups (photon energy 400 eV). The solid line corresponds to a fit with a single component (doublet); (b) N 1s XPS core-level peak of the G-pATP surface. The solid line corresponds to a fit with a single component at 400.09 eV, (c) 50 nm × 50 nm STM image of a SiC/H/G-functionalized surface after incorporating pATP. Imaging conditions: $V = -339$ mV, $I = 0.181$ nA. Inset: 5 nm × 5 nm. Zoom of one molecule taken at $V = 600$ mV, $I = 0.102$ nA; (d) representative Raman spectra of epitaxially-grown graphene on the 4H−SiC surface before (upper trace, red curve) and after (lower trace) incorporation of pATP at the vacancies, showing graphene band assignments. (e) Top: schematic representation of 2,2′ dithiodipyridine (DTP) reaction with thiol groups. Bottom: UV−vis spectrum of the product reaction obtained between pATP-functionalized graphene and DTP. (f) Magnetic field dependence of the Hall resistance channel for the pristine graphene (black dots) and the freshly functionalized G-pATP (red dots), corresponding to our setup to transport by holes and electrons. (g) Top (inset) and perspective views of a computed 3D density-charge redistribution function isosurface (with a positive value of $|\rho_{\text{redist}}(r)| = +10^{-4}$ e⁻ a.u.⁻³) for the functionalized system. The density-charge redistribution function is defined as $\rho_{\text{redist}}(r) = \rho_{\text{tot}}(r) - [\rho_{\text{gr}}(r) + \rho_{\text{mol}(-2H)}(r)]$, where $\rho_{\text{tot}}(r)$ is the total electronic spatial density charge for the whole, once formed, molecule(−2H)@graphene system, and $\rho_{\text{gr}}(r)$ and $\rho_{\text{mol}(-2H)}(r)$ are the total electronic spatial density charges for the graphene with the single atomic vacancy and the pATP(−2H) molecule in the geometric configuration they adopt in the whole system, respectively.

Figure 2d presents a representative Raman difference spectrum recorded from a flat graphene terrace of graphene on 4H−SiC, where the second-order modes of SiC[31] have been subtracted to reveal the characteristic graphene bands. In the upper trace, which represents the graphene surface prior to functionalization, very little evidence for defect-induced modes was found, confirming the high quality of the as-grown graphene layer, where G, 2D, and 2D′ bands appear at $\Delta\nu = 1585$, 2680, and 3232 cm⁻¹, respectively, and the 2D/G ratio of ∼3.2 correlates very well with quasi-free standing monolayer graphene.[32,33] In the lower trace, which corresponds to a pATP-functionalized surface, the defect-induced bands D and D + D′ clearly emerge at around 1362 and 2962 cm⁻¹, respectively, because of the included N atom of the pATP that is sp³ coordinated inducing a distortion of the graphene sheet. However, the 2D/G ratio remains little changed at >3, attesting the structural integrity of the graphene sheet, and the incorporation of the N atom into the basal plane is an essential part of this process.

In order to assess the effect of functionalization on the electrical properties, the pristine graphene and G-pATP surfaces were compared via a magnetotransport method.[34–36] The results indicate that the density of carriers does change because of the charge redistribution after bonding and mobility is reduced by a 24% (Figure 2f). At room temperature, there is a striking change in the slope of $R_{xy}$ resistance versus magnetic field when data from the pristine sample are compared with data taken immediately after the production of G-pATP (Figure 2f). This change in the slope plot at low fields (±1 T) corresponds to a change in the sign of the charge carriers from holes (with an estimated density of $5.2 \times 10^{14}$ cm$^{-2}$) in the pristine sample to electrons (with a density of $1.3 \times 10^{14}$ cm$^{-2}$) in the functionalized sample. These experiments were performed on an epitaxially grown single-layer graphene surface; therefore, the sign of doping is as expected. The effects of functionalization are also visible in the highest level of noise apparent in the Hall channel for the functionalized sample data. There is also a decrease in the estimated mobility (2390 cm$^2$ V$^{-1}$ s$^{-1}$ in the pristine sample and 1600 cm$^2$ V$^{-1}$ s$^{-1}$ after functionalization). Further, the G-pATP sample was measured again after 1 year storage under ambient conditions, and the electrical properties, in terms of charge carriers and density, were essentially preserved.

At such low surface coverage, neither Raman nor infrared spectroscopy could detect the thiol functionalities; thus, their presence was assayed in a liquid environment by dipping a functionalized graphene surface for 1 h into a DTP solution.[37] The free thiol groups react with DTP forming 2,2′-dipyridyl disulfide and releasing 2-mercaptopyridine that quickly tautomerizes into 2-thiopyridone (top part of Figure 2e). The formation of these compounds indicates the presence of free thiols on the graphene surface giving a UV absorption maximum at $\lambda = 343$ nm (molar extinction coefficient at 343 nm = 8080 M$^{-1}$ cm$^{-1}$), as shown in Figure 2e. The peak observed at 269 nm is associated with unreacted DTP.[38] Thus, we can confirm that the pATP molecules are covalently anchored through the amine group to the graphene surface exposing free thiol groups, which are stable in air and available for subsequent reactions.

Figure 2g presents the computed 3D isosurface for a positive value of $+10^{-4}$ e a.u.$^{-3}$ of the density-charge redistribution function (defined in the figure caption) for the G-pATP system and shows how the incorporation of the pATP molecule, via the doubly dehydrogenated N atom within a single atom vacancy (SAV) of the graphene lattice, induces a strong electronic charge spatial localization in the surroundings of the incorporated N atom in a third-order symmetrical way (see the circular inset in Figure 2g), which has been quantified by a Bader population analysis to comprise 0.96 e$^-$. This electronic charge accumulation comes from both the rest of the anchored molecule (around 0.5 e$^-$) and from charge depletion in graphene regions outside the functionalization centers (around 0.56 e$^-$). Interestingly, the electronic charge depletion and redistribution within the rest of the molecule breaks down the aromatic resonance of the ring, and some electronic charge seems to accumulate in the S atom of the thiol group. This charge accumulation potentially enhances the reactivity of this group, thus positively favoring reaction at the SH site of the anchored molecule. The effect of the solvation has also been studied by DFT calculations. We have repeated some calculations having into account the water solvation influence by the polarizable continuum model (PCM) within the default self-consistent reaction field method as implemented in Gaussian09.[39] For this purpose, we have computed again the total energy of the full molecule ($-2$H)@graphene system and of the graphene with the single atomic vacancy and the pATP($-2$H) molecule subsystems by separate within the PCM model under water solvent conditions. The result of the calculations yields an anchoring energy of 11.82 eV, to be compared with the value of 12.21 eV without accounting the water solvation environment. This small energy difference does not alter substantially the robustness scenario of the bonding after the functionalization.

Once confirmed the robustness of the thiol-functionalized graphene upon exposure to both air and liquid environments, we focus on its capacity for conjugation to metal NPs or biomolecules.[40]

In order to couple metal NPs, we take advantage of the strong and well-studied S—metal bond that forms spontaneously and allows to strongly and easily link any noble metal NP to the functionalized graphene surface.[41] For the technological use of graphene/NP assemblies, the NPs need to be securely anchored to a particular surface. The versatility of the present methodology is well suited to this purpose, and many different types of NPs can be robustly coupled to graphene. To demonstrate this, two types of Au NPs of different origins were employed, described in Figure 3: NPs produced in a gold-salt solution and capped with citrate and NPs produced using a multiple ion cluster source (MICS) in UHV.[42,43]

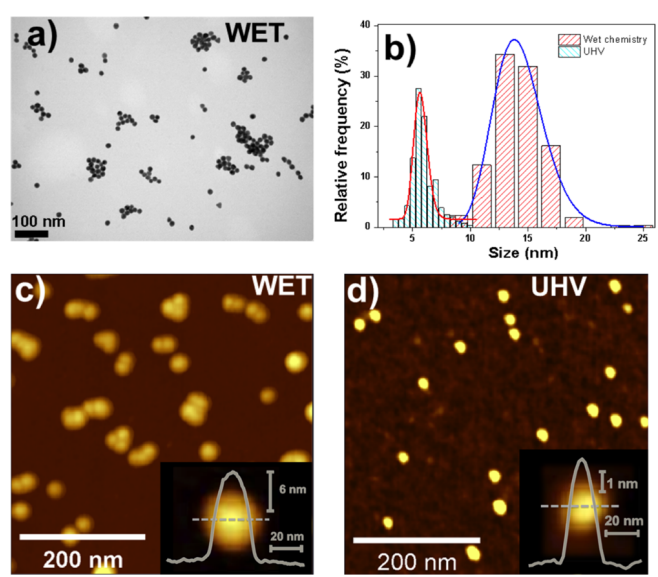

**Figure 3.** (a) TEM micrograph showing Au NPs prepared by wet chemistry; (b) size distributions, with the corresponding log-normal fits, of Au NPs prepared under wet chemistry and UHV conditions obtained from the heights of the corresponding dynamic-mode AFM images; (500 nm × 500 nm) of the Au NPs deposited on graphene via (c) wet chemistry and (d) under UHV conditions. Insets correspond to characteristic surface profiles of Au NPs. Note that the lateral sizes of the NPs are increased because of tip convolution effects.

Both transmission electron microscopy (TEM) (Figure 3a) and atomic force microscopy (AFM) images (Figure 3c) of particles prepared by wet chemistry methods show that these have a tendency to aggregate, whereas those prepared in UHV (Figure 3d) remain completely isolated, with a low diffusion on this substrate. In the latter case, no organic shell or surfactants are required to avoid NP aggregation, which may prove advantageous for some plasmonic or optoelectronic applications.

The particle size and size distribution also vary for both types of NPs, as obtained from the AFM height measurements, undertaken in order to avoid tip convolution effects that coarsen the lateral image size. After measuring more than 50 Au NPs from each sample type, the size distribution histogram in Figure 3b was obtained. An average NP size of 5.7 ± 0.6 nm was found for Au NPs prepared under UHV conditions and 14 ± 2 nm for those prepared by wet chemistry.

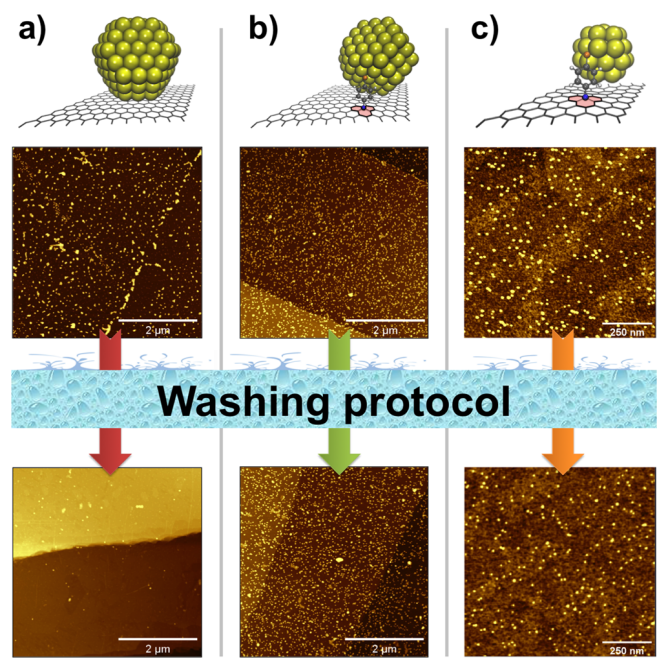

**Figure 4.** AFM images of Au NPs deposited on (a) as-received SiC/H/G substrate, (b) SiC/H/G-pATP substrate following the wet chemical route, and (c) SiC/H/G-pATP substrate following the UHV physical route (see text). Top row: as-deposited; Bottom row: after 1 washing cycle. Note that the image sizes in (c) are smaller because of the lower Au NP size.

With regard to the density of NPs deposited, the top row of Figure 4 presents large-scale AFM images for three different cases: Au NPs deposited on pristine, nonfunctionalized graphene (column A), and Au NPs deposited on G-pATP surfaces, following either the wet chemical (column B) or the UHV physical (column C) route. The figure shows that more NPs are seen in the functionalized surface, with $1.2 \times 10^{10}$ cm$^{-2}$ and $3 \times 10^{10}$ cm$^{-2}$ for chemical and physical deposition, respectively, than that for the pristine G/Au system, with $4 \times 10^{9}$ cm$^{-2}$, suggesting that the deposition of Au NPs is relatively enhanced by the presence of pATP.

In most studies of decoration of high-quality graphene with NPs, these are simply deposited on the surface from solutions,[44,45] although a recent study shows −COO−-modified Au NPs electrostatically linked to graphene surfaces functionalized via thiol−ene click chemistry, but NP stability or resistance was not addressed.[29,46] To test the binding strength of the Au NPs to the G surfaces, the respective samples are subjected to a washing cycle, the corresponding images being displayed in the bottom row of Figure 4. Clearly, in the case of the pristine G surface, almost all of the NPs were removed from the surface. However, for the G-pATP surfaces, the density of NPs does not appear to change, within the statistical error, after washing for both chemical and physical routes. This is due to the high strength of the covalent link between the thiol group and the Au NPs.

Furthermore, an important advantage of our methodology is that the number of available thiol moieties on the surface can be relatively easily tuned up to a maximum estimated density of approximately $8 \times 10^{12}$ cm$^{-2}$, which would correspond to approximately 0.1−1% thiol-functionalized surface.[11] Thus, it is easy to control the typical average separation between immobilized NPs. Finally, it is important to point out that we have successfully linked Au NPs to graphene surfaces that were functionalized 5 months earlier and stored under ambient conditions, showing a high resistance of the G-pATP surface to ageing.

With regard to the development of biohybrid interfaces, we have also explored the suitability of the G-pATP surfaces to immobilize an in vitro selected thiol-derived ssDNA aptamer with high specificity for the recognition of a target protein, where we explored the possibility of forming a stable disulfide bridge bond between the S atom of the free thiol group of the pATP molecule linked to the graphene surface and a terminal S atom present in the thiol-modified aptamer molecule. In this respect, we employed a previously obtained 76 nucleotides-long ssDNA aptamer with selectivity for the protein PCBP-2, a pleiotropic protein that participates in a number of cellular processes including transcriptional and translational regulation. We use the protocol described in the experimental part to first produce an aptamer-modified graphene surface and, second, to expose it to a solution containing target protein molecules. The final dried surfaces are studied in air using AFM, and a typical AFM image of the G-pATP/aptamer/PCBP-2 system is shown in Figure 5. Here, two different types of rounded and compact structures were observed: some small ones with sizes in the 16−20 nm range and a more reduced number of large protuberances with lateral sizes in the 28−32 nm range, albeit the lateral sizes of all of these features may be coarsened by tip convolution effects as well as by the tip load. The different heights observed in the surface profiles corresponding to the path 1-2-3 in the AFM image are shown in Figure 5b. Clearly, the smaller structures show heights in the 0.5−1.5 nm range and likely correspond to free aptamers bound to the G-pATP surface, whereas the larger structures reach heights of 3 nm and can be assigned to aptamer/PCBP-2 complexes. Analysis of the heights of 400 features (not shown) like those shown in Figure 5a presented two maxima at around 0.9 and 3.1 nm, corresponding to aptamer and aptamer/protein structures, respectively. These data are consistent with our experimental conditions in which an aptamer/protein ratio of 10:1 was employed, and only a limited fraction of the aptamers is expected to be bound to the target PCBP-2.

It should be pointed out that one of the added values of using a biosensing system based on aptamers instead of antibodies is their reversibility with temperature.[22] The denaturing temperature used for our aptamer (with a length of 76 nucleotides) without compromising its integrity is 95 °C. Unlike antibodies, which undergo irreversible denaturation above 37 °C, aptamers can refold into their original, active conformations when optimal temperature (in our case, 25 °C) is restored. Therefore, the aptamer is fully denatured at 95 °C, and the target molecule (in this case, the protein PCBP-2) is completely released, and then the surface immobilized aptamers can be reused. The G-pATP system is perfectly capable of resisting this temperature level, which allows repeated reutilization of the same G-pATP/aptamer sensing platform. It is important to note that the developed chemistry in this work is perfectly capable of resisting this denaturation temperature.

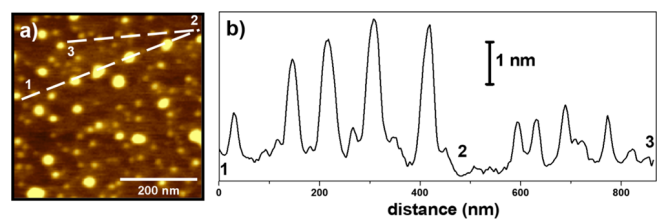

**Figure 5.** (a) 500 nm × 500 nm AFM image taken in dynamic mode in air of G-pATP/aptamer/PCBP-2, after a washing step with selection buffer (SB). The line profile connecting points 1, 2, and 3 in the image is shown in panel (b). The image displays 0.5−1 nm-height and 16−20 nm-wide features (that can be assigned to free aptamers), as well as a smaller number of 1.5−3 nm-height and 28−32 nm-wide features (corresponding to aptamers bound to the target protein).

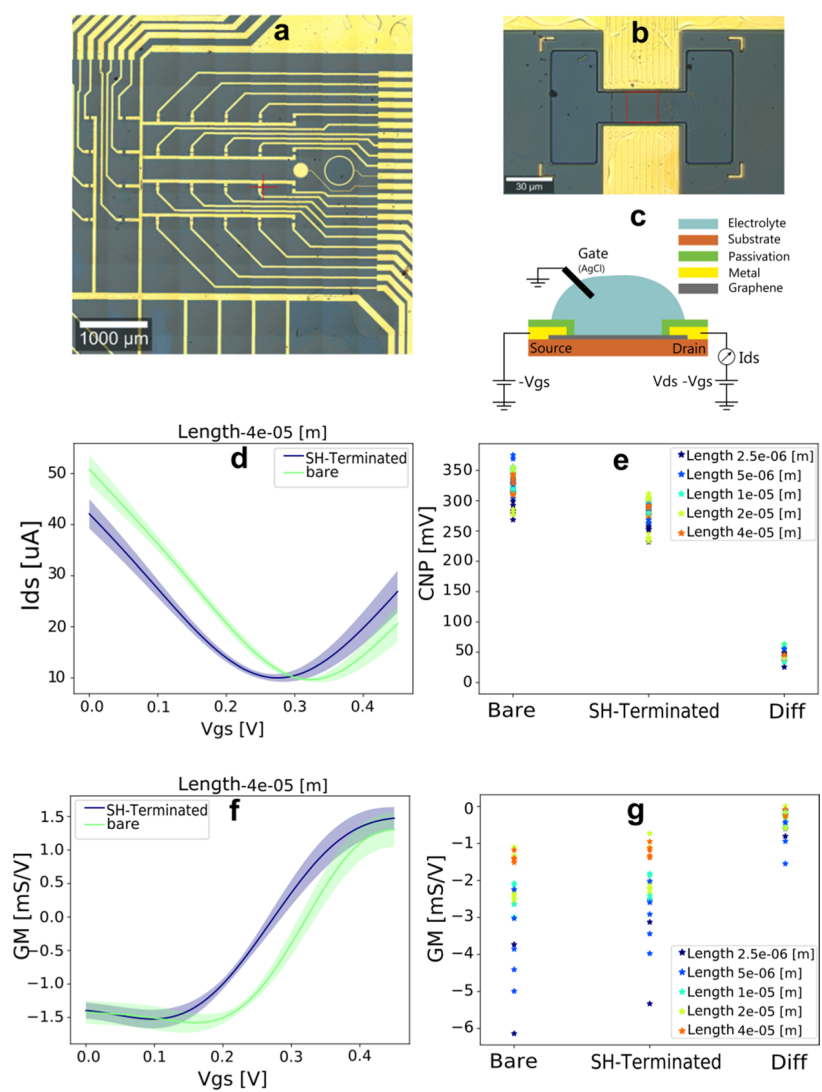

**Figure 6.** (a) Optical microscopy image of the solution-gated field-effect transistor (gSGFET) device with the graphene transistors displayed (red cross indicates a transistor), (b) Transistor for a 40 × 20 μm channel size (red square on the graphene channel) and (c) cross section of the gSGFET device. The gate−source polarization is applied on the gate, and the drain−source polarization is applied on the drain. The $I_{ds}$ is measured on the drain side. (d−g) Characterization curves of the CVD transistor before and after functionalization with pATP. (d) $I−V$ transfer curves for 40 × 40 μm transistor channel size, (e) CNP values at different channel lengths before and after functionalization and the difference between both, (f) gm curves for 40 × 40 μm channel size, and (g) gm values at different channel lengths before and after functionalization and the difference between both.

The binding mechanism is different in both cases. In the case of metal NPs, they are based on the strong affinity of the sulfur-containing compounds for noble metal surfaces, specifically gold. The mechanism of binding NPs to the functionalized G surface consists of a reductive elimination of hydrogen of the S−H bond by the metal surface. However, in the case of G-pATP functionalization with the thiolated aptamer, it relies on the oxidation of both sulfhydryl groups reaching a disulfide bridge.[47] Gold-thiolate and disulfide bond cleavage occur via reduction. Hence, the solvent would only affect if it contains reductant agents, which is not our case. Additionally, because basic pH could also promote gold-thiolate dissociation, we employed pH 7.4 for the functionalization process.[48] Finally, the strong affinity between gold and the thiol group and the S−S bond strength guarantee the robustness of the functionalization and binding processes, leading to the stability of the different platforms developed in this work under the different experimental conditions.[11]

## 2.1. Perspectives, a Platform for gFET.
As it has been shown, the presented methodology exhibits potential applicability in two systems of broad interest, namely, metal NPs and (bio)sensing. The use of gFET as sensing devices has shown certain advantages in biomedical applications, such as described for PoC (point of care) biosensing of biomolecules[49] or electrical brain recording.[50] However, for using the gFET as biosensors, the graphene channel has to be chemically modified to graft bioreceptors that will detect the specific biomarkers. In this specific application,[50] keeping the electrical performance after the functionalization procedures is really important to ambition dual sensing applications, recording and sensing biomolecules in parallel.[51]

In order to advance toward the implementation of these functionalized graphene surfaces into devices showing electrical read out for (bio)sensing applications, the impact of this controlled surface functionalization methodology on the electrical properties of graphene needs to be evaluated. For this purpose, we have fabricated and characterized prototypical graphene solution-gated field-effect transistors (gSGFET),[52] which were measured before and after functionalization with pATP.

Figure 6a shows an optical image of a device consisting of a 25-transistor array, where a series of single-layer chemical vapor deposition (CVD)-graphene channels are connected by two metallic contacts, the drain and the source, with a reference electrode (Ag/AgCl) immersed in an electrolyte solution, which is used as the gate terminal that modulates the conductivity of the graphene channel. The image in Figure 6b shows a close up of one of the graphene channels and the gold contacts of the transistor corresponding to the area marked by a red cross in Figure 6a. A schematic representation of the device is shown in Figure 6c. The standard electrical characterization is derived from the gSGFET transfer curve (Figure 6d), where the drain−source current, $I_{ds}$, is obtained as a function of the gate−source voltage, $V_{gs}$ (−0.2−0.6 V), for a fixed drain−source voltage, $V_{ds}$ (0.1 V). From this curve, both the charge neutrality point (CNP), corresponding to the voltage at the minimum Ids current and minimum of the curve, and the transconductance (gm), corresponding to the slope of the $I-V$ curve, can be evaluated. The latter, which is the derivative of the $I_{ds}$ versus $V_{gs}$ (Figure 6d), is related to the geometrical parameters of the transistor (width to length ratio) and the graphene charge mobility. The CNP can be related to the doping of the graphene: a p-doped surface shifts the CNP to more positive values of $V_{gs}$, whereas an n-doped surface shifts to more negative ones.[53] This doping arises principally from the nature of the molecules attached to the surface and the charges trapped near the graphene lattice. The transfer curves ($I_{ds}$ vs $V_{gs}$) obtained before and after functionalization for a 40 $\mu$m long gSGFET are shown in Figure 6d, where it can be observed that the functionalization mainly affects the CNP, shifting it to lower potentials, whereas the other parameters remain constant. In Figures 6e−g, the values of CNP and gm obtained before and after functionalization are shown, as well as the variation evaluated for each transistor before and after functionalization. The average variation in CNP observed over the 25 transistors is 45.7 ± 9.8 mV, suggesting that the molecular coverage on the surface does not induce a large doping in the graphene lattice. The broad dispersion in gm observed in the measurements before and after functionalization is due to the different channel lengths of the transistors. From the results obtained, the gm after functionalization is slightly reduced to a nonrelevant value of 0.401 ± 0.340 m S, suggesting that the charge mobility in the graphene lattice is unaffected by the process. Indeed, the small gm reduction could be attributed to variations in the coupling capacitance, although further evaluations need to be undertaken to fully assess this behavior. Thus, it can be concluded that is a valid functionalization strategy to tailor the CVD graphene surface in a device with thiol-terminated moieties, and it opens the door to versatile thiol chemistry to a broad range of biosensing experiments using this type of devices.

## 3. CONCLUSIONS

In summary, we present here a facile and controllable methodology to decorate a graphene surface with covalently linked organic spacers containing exposed thiol functionalities that can be employed in diverse chemical linking strategies. In particular, we illustrate the linking of metal NPs and aptamers, bisensor-relevant biomolecules, to a graphene surface. This methodology opens the door to the growth of controllable and stable nanobiohybrid structures on graphene platforms, with a broad applicability in plasmonic, biotechnology, and biomedicine.

## 4. EXPERIMENTAL METHODS

### 4.1. Graphene Surface Functionalization.
Although we have employed several types of graphene surfaces, all with similar results, most of the experiments reported were performed on high-quality graphene layers epitaxially grown on 4H−SiC(0001) by CVD with or without H under the superficial graphene layer. These surfaces behave as quasi free-standing monolayer graphene, with Hall mobility values above 8 000 $cm^2$ $V^{-1}$ $s^{-1}$.[54] On the other hand, the immobilization protocol has also been performed directly on devices, where CVD graphene sheets were obtained by electrochemical delamination and transferred onto $SiO_2$ wafers.

The main steps in the functionalization process are illustrated in Figure 1. The strategy used to covalently link thiol moieties to the graphene basal plane was adapted from a recently reported protocol used to couple para-aminophenol molecules.[11] It begins in the UHV chamber by creating a network of single atomic vacancies on the surface via gentle $Ar^+$ sputtering of the graphene surface (Figure 1, step 1).[55,56] The functionalization experiments were performed in situ in UHV with a base pressure below $2 \times 10^{-10}$ mbar. In the first step, the substrate was degassed, followed by an annealing cycle at 250 °C for 15 min to remove any physiosorbed contamination from the surface. Vacancies were created in the graphene lattice by bombarding with argon ions ($Ar^+$) for 90 s using an electron-impact ion gun. The acceleration energy of the ions inside the ion gun was 140 eV, the sample current was 1 $\mu$A, and the gas pressure during this process was maintained at $1 \times 10^{-7}$ mbar. The sample was then annealed at 500 °C for 5 min. Before dosing pATP molecules (Sigma-Aldrich, purity 99%), these were purified by pumping with a turbomolecular pump ($P = 10^{-8}$ mbars) for >6 h to eliminate any impurities. Subsequently, the ion-treated graphene surface was exposed to about 10 Langmuir of pATP (1 L = $10^{-7}$ mbar s). The G-pATP surfaces were either maintained in UHV for further in situ reaction with Au NPs or removed from the chamber and stored under ambient conditions for ex situ reaction with Au NPs or aptamers.

### 4.2. NP Fabrication.
Citrate-stabilized gold NPs were synthesized according to the Turkevich method.[57] The Au content in the samples was determined by inductively coupled plasma-optical emission spectrometry (ICP-OES, PerkinElmer

Optima 2100 DVICP) after digestion of the Au NP suspension with a mixture of $HNO_3$:HCl (1:3 ratio v/v) and dilution with ultrapure water. Graphene (1 cm$^2$) functionalized with pATP was placed into a small plastic vial containing 2.5 mL of the aqueous suspension (0.1 mg Au/mL) of Au NPs and left, without agitation, for 15 h. Next, the graphene sample was introduced into a Petri dish in the presence of 5 mL of water and stirred in an orbital shaker for 15 min. Subsequently, in order to eliminate any residues of unconjugated Au NPs, the graphene surface was thoroughly washed with ultrapure water. Finally, the washed sample was dried in an oven at 50 °C for 2 h. The same procedure was applied to pristine, non-functionalized graphene. The NPs on the graphene surfaces were observed by TEM and AFM.

Au NPs were fabricated in UHV via a gas-phase synthesis route using a MICS,[43] equipped with three 1″ magnetrons. In this case, one of the magnetrons was loaded with an Au target (99.99% purity). To ensure minimal contamination, all gas pipes used to inject Ar (95.5% purity) are made of stainless steel and vacuum-sealed. The MICS is connected to the main UHV chamber, where the substrate is located, through a small orifice with a nozzle. The distance of the sample to the MICS exit during deposition was 200 mm. The working parameters employed in the MICS for Au NP fabrication were a power of 8 W; an aggregation distance to the exit slit of 70 mm; and total Ar flux of 100 sccm with an Ar flux through the Au magnetron of 30 sccm. Once formed, the Au NPs travel through the vacuum to the main chamber where they soft land on the surface of the substrate, ensuring the absence of deformation of the Au NPs due to their low energy.[58] The base pressure in the system was below $5 \times 10^{-10}$ mbar.

### 4.3. Binding of Aptamers to G-pATP and Target Protein Recognition.

Aptamers are RNA or ssDNA oligonucleotides that are selected in vitro from a vast library of synthetic oligonucleotides (in general, from $10^{13}$ to $10^{15}$) with random sequence[21] using an amplification-selection method termed "systematic evolution of ligands by exponential enrichment" or SELEX,[59] which can bind with high affinity and specificity to a given target molecule. Aptamers possess a specific three-dimensional structure in solution that depends on their sequence and on the physicochemical features of the folding buffer, including temperature, pH, ionic strength, and concentration of divalent cations.

We have used a 76 nucleotides-long ssDNA aptamer specific to the protein PCBP-2 (also known as hnRNP E2 or CP-2), which is a member of the cellular heterogeneous nuclear ribonucleoprotein family that mediates relevant biological processes including mRNA stabilization, transcriptional regulation, translational control, and apoptotic program activation.[60–62] Among the PCBP-2-specific aptamers previously obtained and analyzed by means of a variant of the SELEX process,[63] we selected the ssDNA aptamer with the highest affinity for the target protein, with a dissociation constant $K_d$ of 8.4 nM. It is termed 05DS10-21, and its nucleotide sequence (5′ thiol-modified, purchased from IBA GmbH, Göttingen, Germany) is 5′-GCGGATCCA-GACTGGTGTGGAGGTTAGCCGAAACACGTA-TACGCGTATTTATCCTCGGGCCCTAAAGA-CAAGCTTC-3′. The recombinant protein PCBP-2 used as the target in this work (purchased from Abnova, Taipei, Taiwan) is a 362 amino acids-long globular protein that includes a GST-tag at its N-terminal end (MW of 65.56 kDa).

The 5′ thiol group of the aptamer was introduced to establish a disulfide bond with the free thiol group of the G-pATP. The terminal sulfur atoms of thiolated DNA have a tendency to form dimers in solution, especially in the presence of oxygen. Therefore, dithiothreitol (DTT) was used as a reducing or "deprotecting" agent for thiolated DNA. Typically, a 10 mM DTT solution in 100 mM sodium phosphate buffer (pH 8.3) was mixed with a 100 $\mu$M thiolated aptamer solution and allowed to react for 1 h at room temperature. Then, DTT was removed by filtration using a 3 K Amicon Ultracentrifugal filter. Subsequently, 30 $\mu$L of the thiolated ssDNA aptamer (at 1 $\mu$M concentration in selection buffer, SB, composed of 100 mM NaCl, 6 mM $MgCl_2$ and 100 mM HEPES pH 7.4) was renatured (by incubation at 95 °C for 10 min, and then at 37 °C for 10 min). The renatured, thiolated ssDNA was deposited onto the G-pATP surface and left to react via disulfide bridge[47,64,65] in a humidity chamber for 20 min at controlled temperature (25 °C). Then, the G-pATP/aptamer surface was gently washed with ultrapure, DEPC-treated milliQ water, and finally air-dried. The aptamer-functionalized graphene samples were subsequently incubated with 30 $\mu$L of a 100 nM solution (in SB) of recombinant PCPB-2 protein for 20 min, in a humidity chamber at 25 °C. Then, the sample was gently washed with SB to discard the protein molecules that could have nonspecifically bound to the surface. Finally, the G-pATP/aptamer/protein samples were AFM imaged in air, as previously described.[66]

### 4.4. Characterization Techniques.

XPS was undertaken at the PEARL Beamline of the Swiss Light Source. The beamline produces a photon flux of $2 \times 10^{11}$ ph/s at 1 keV on sample. XPS spectra were fitted with an Igor Pro macro using Voigt (Lorentzian and Gaussian) curve profiles. In all cases, a Shirley-type background was subtracted from the raw data.

Raman spectra were obtained in the Raman laboratory of the characterization service of the ICTP–CSIC employing a Renishaw inVia Reflex-dispersive Raman spectrometer incorporating a Leica microscope, using a 100× (N.A. = 0.85) objective with a 514.5 nm (2.41 eV) laser power of approx. 2 mW at the sample surface, and multiple scans in order to obtain adequate signal-to-noise ratios.

The in-air AFM measurements were performed in the dynamic mode with a Nanoscope IIIA (Veeco) system, an Agilent 5500 PicoPlus, and a Cervantes system from Nanotec Electronica S.L. under ambient conditions. We employed silicon tips with nominal force constants of 40 N/m and softer silicon cantilevers of 2–4 N/m for imaging the aptamer/protein complexes.

TEM measurements were performed using a JEOL JEM 1011 with a Gatan ES1000Ww camera. TEM samples were prepared by placing one drop of a dilute suspension of Au NPs in water on a carbon-coated copper grid, and the solvent was slowly evaporated at room temperature.

Magnetotransport was investigated over macroscopic dimensions (10 mm × 10 mm) in the square van der Pauw configuration using a Quantum Design 9T PPMS (R) and press-on contacts from Wimbush Scientific (R). Resistance measurements were collected in helium atmosphere (5 Torr) along the sample edges ($xx$ and $yy$) and diagonal ($xy$) directions.

### 4.5. Theoretical Computational Methods.

Physical and electronic properties of the pATP molecules incorporated within a SAV created in the graphene were using a set of accurate DFT-based calculations by the plane-wave scheme implemented in the QUANTUM ESPRESSO simulation package.[67] All calculations accounted for dispersion forces within the DFT + D approach via an empirical efficient vdW R-6 correction.[68,69] The revised version of the generalized gradient-corrected approximation of Perdew, Burke, and Ernzerhof (rPBE) has been used to account for the exchange−correlation effects,[70] and Kohn−Sham equations were solved using a periodic supercell geometry. Rabe−Rappe−Kaxiras−Joannopoulos ultrasoft pseudopotentials[71] have been adopted to model the ion−electron interaction in the H, C, N, and S atoms. The Brillouin zone was sampled by employing an optimal $[4 \times 4 \times 1]$ Monkhorst−Pack grid.[72] One-electron wave functions were expanded on a basis of planewaves with energy cutoffs of 450 and 550 eV for the kinetic energy and the electronic density, respectively, which have been adjusted to achieve sufficient accuracy to guarantee a full convergence in total energy and density.

**4.6. gSFET Device Fabrication and Testing.** CVD graphene transistors were fabricated using graphene layers grown by a CVD process using a 25 $\mu$m thick copper foil of 99.8% metal basis obtained from Alfa Aesar. The copper foils were cleaned with acetic acid, then rinsed with deionized (DI) water, followed by cleaning cycles with acetone, DI water, and isopropyl alcohol. The sample dimensions were 6 × 5 cm², and the growth conditions were 10 min at 750 °C, 2 sccm $H_2$ and 5 min at 800 °C, 25 sccm $CH_4/H_2$. A 700 nm poly(methyl methacrylate) (PMMA) layer was deposited on top of the graphene surface. The graphene layer was separated from copper using an electrochemical delamination process following the procedure described by Rosa et al.[73] without the use of a frame. After transfer, the layers were baked at 180 °C for 2 min and the PMMA was subsequently removed by rinsing with acetone and isopropanol.

A Si/SiO₂ wafer was used as a substrate. The first layer of Ti/Au metal contacts was deposited by electron-beam vapor deposition and subsequently structured by optical lithography. Then, CVD graphene was transferred to the wafer using the process described in the previous section. The graphene active area W = 40 $\mu$m × L = 2.5/5/10/20/40 $\mu$m of the sensors was then defined by oxygen plasma in a reactive ion etching system. A second metallization layer of Ni/Au was then evaporated and lithographically defined, followed by a lift-off step. To electrically insulate the device, a 2 $\mu$m thick SU-8 epoxy photoresist (SU-8 2005 Microchem) layer was spin-coated on top and defined in such a way that only the graphene area was left uncovered.[52] The chips were finally individually cut.

The electrical characterization was performed using a PA Suss wafer prober with an Agilent 41000 semiconductor analyzer. The graphene transistors were in contact with an electrolyte solution of 0.1 M KCl in phosphate-buffered saline solution. A reference electrode Ag/AgCl was used as a gate.

■ **AUTHOR INFORMATION**

**Corresponding Author**
*E-mail: gago@icmm.csic.es.

**ORCID** ⦿
Carlos Sánchez-Sánchez: 0000-0001-8644-3766
Jose A. Garrido: 0000-0001-5621-1067
María del Puerto Morales: 0000-0002-7290-7029
José A. Martín-Gago: 0000-0003-2663-491X

**Author Contributions**
Rebeca Bueno and Marzia Marciello contributed equally to this work. The manuscript was written through contributions of all authors. All authors have given approval to the final version of the manuscript.

**Notes**
The authors declare no competing financial interest.

■ **ACKNOWLEDGMENTS**

This work was supported by the European Union's Horizon 2020 research and innovation programme under grant agreement No 696656 (Graphene Flagship-core 1) and no 785219 (Graphene Flagship −core 2); UE FP7 ideas: ERC (grant ERC-2013-SYG-610256 Nanocosmos) and Spanish MINECO grants MAT2014-54231-C4-1-P, MAT2014-54231-C4-4-P, MAT2017-85089-C2-1-R, MAT2014-59772-C2-2-P, and BIO2016-79618-R (funded by EU under the FEDER programme), as well as the Nanoavansens program from the Community of Madrid (S2013/MIT-3029). This work has made use of the Spanish ICTS Network MICRO-NANOFABS partially supported by MINECO and also the ICTS NANBIOSIS, more specifically the Micro-Nano Technology Unit of the CIBER in Bioengineering, Biomaterials & Nanomedicine (CIBER-BBN) at the IMB-CNM. We are grateful to Matthias Muntwiler for his assistance with experiments in the PEARL beamline in the SLS facility. Finally, we acknowledge the TEM and ICP services at the CNB and ICMM institutes, respectively. CSS acknowledges the MINECO for a Juan de la Cierva Incorporación grant (IJCI-2014-19291). M. Marciello is grateful to the Comunidad de Madrid (CM) and European Social Fund (ESF) for supporting her research work through the I+D Collaborative Programme in Biomedicine NIETO-CM (B2017-BMD3731).